\documentclass[10pt,twoside,final,twocolumn,a4paper,notitlepage]{article}
\pdfoutput=1

\usepackage[utf8]{inputenc}
\usepackage{amssymb}
\usepackage{program}
\usepackage{ifthen}
\usepackage[dotinlabels]{titletoc}
\usepackage[pdftex]{graphicx}
\usepackage{moreverb}
\usepackage{multirow}
\usepackage{multicol}
\usepackage{url}
\usepackage{listings}
\usepackage{float}
\usepackage{paralist}
\usepackage[font=footnotesize]{subfig}
\usepackage{booktabs}
\usepackage{flushend} 
\usepackage[scale=.87]{geometry}
\usepackage[pdftex]{hyperref}  
\usepackage{titling}

\usepackage[T1]{fontenc}
\usepackage[scaled=.8]{beramono}   
\usepackage{scalefnt}              

\usepackage{color}
\definecolor{codecolor}{rgb}{0,0,0.3}
\definecolor{gray97}{gray}{.97}
\definecolor{gray45}{gray}{.45}

\graphicspath{{./Plots/}{./}}

\newcommand{\fixedwidthfigure}[2]{
  \begin{figure}[htbp!]
    \begin{center}
      \includegraphics[width=.98\linewidth]{#2}
      \caption{#1}
      \label{fig:#2}
    \end{center}
  \end{figure}
}

\newcommand{\fixedscalefigure}[2]{
  \begin{figure*}[htbp!]
    \begin{center}
      \includegraphics[scale=.7]{#2}
      \caption{#1}
      \label{fig:#2}
    \end{center}
  \end{figure*}
}

\newcommand{\rnaseq}{\protect\resizebox{!}{1.7ex}{\sf HPG-aligner}}

\makeatletter
\newcommand{\makefirstpagetitle}{%
  \begin{titlepage}
    \begin{center}
      
      \sffamily
      
      \null

      \vfill
      
      \includegraphics[width=3cm]{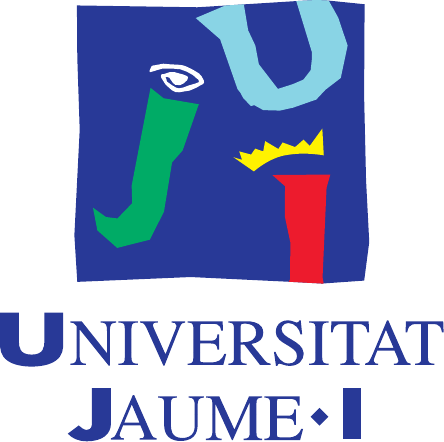}
      
      \bigskip{}

      {\large Technical Report ICC~2013-03-01}

      \vfill

      {\huge \textbf{\thetitle}\par}
      
      \bigskip
      \bigskip
      \bigskip
      
      {\Large \theauthor\par}

      \bigskip
      \bigskip
      
      \textsuperscript{*}%
      Dpto. de Ingeniería y Ciencia de los Computadores\\
      Universidad Jaume~I\\
      12006 - Castellón\\
      Spain\\
      {\{martineh,barrachi,castillo,quintana\}@uji.es}

      \bigskip{}

      \textsuperscript{\dag}%
      Computational Genomics Institute\\
      Centro de Investigación Príncipe Felipe\\
      46012 - Valencia\\
      Spain\\
      {\{jtarraga,imedina,jdopazo\}@cipf.es}

      \bigskip
      \bigskip

      {\Large \@date\par}

    \end{center}

    \vfill

    \vfill

    \vfill

  \end{titlepage}

  \thispagestyle{empty}

  \cleardoublepage{}

}
\makeatother

\title{Concurrent and Accurate RNA Sequencing\\ on Multicore Platforms}

\author{%
  Héctor Martínez\textsuperscript{*} \and
  Joaquín Tárraga\textsuperscript{\dag} \and
  Ignacio Medina\textsuperscript{\dag} \and
  Sergio Barrachina\textsuperscript{*} \and
  Maribel~Castillo\textsuperscript{*} \and
  Joaquín Dopazo\textsuperscript{\dag} \and
  Enrique S. Quintana-Ortí\textsuperscript{*}}

\hypersetup{%
 pdftitle={\thetitle{}},
 pdfauthor={\theauthor{}},
  pdfsubject={HPG-Aligner description},
  pdfkeywords={RNA, Short-read alignment, Burrows-Wheeler Transform,
    Smith-Waterman's Algorithm, high performance computing, multicore
    processors.},
  colorlinks=true,
  linkcolor=black,   
  anchorcolor=black, 
  citecolor=black,   
  filecolor=black,   
  menucolor=black,   
  urlcolor=black,    
}

\begin{document}

\makefirstpagetitle{}

\setcounter{page}{1}

\twocolumn[
  \begin{@twocolumnfalse}
    \hfill{} Technical Report ICC~2013-03-01

    \maketitle

    \begin{abstract}

      In this paper we introduce a novel parallel pipeline for fast
      and accurate mapping of RNA sequences on servers equipped with
      multicore processors.  Our software, named
      \rnaseq{}\textsuperscript{1}, leverages the speed of the
      Burrows-Wheeler Transform to map a large number of RNA fragments
      (reads) rapidly, as well as the accuracy of the Smith-Waterman
      algorithm, that is employed to deal with conflictive reads.  The
      aligner is complemented with a careful strategy to detect splice
      junctions based on the division of RNA reads into short segments
      (or seeds), which are then mapped onto a number of candidate
      alignment locations, providing useful information for the
      successful alignment of the complete reads.

      Experimental results on platforms with AMD and Intel multicore
      processors report the remarkable parallel performance of \rnaseq{},
      on short and long RNA reads, which excels in both execution time and
      sensitivity to an state-of-the-art aligner such as TopHat~2 built on
      top of Bowtie and Bowtie~2.


    \end{abstract}

    \bigskip{}

  \end{@twocolumnfalse}
]

\renewcommand{\thefootnote}{\fnsymbol{footnote}}
\footnotetext[1]{Dpto. de Ingeniería y Ciencia de los Computadores,
  Universidad Jaume I,
  12006 - Castellón, Spain.\\
  {\{martineh,barrachi,castillo,quintana\}@uji.es}}
\footnotetext[2]{Computational Genomics Institute, 
  Centro de Investigación Príncipe Felipe, 46012 - Valencia, Spain.\\
  {\{jtarraga,imedina,jdopazo\}@cipf.es}}

\renewcommand{\thefootnote}{\arabic{footnote}}

\setcounter{footnote}{1}

\footnotetext{\rnaseq{} is an open-source software application. It can
  be downloaded from \url{http://www.opencb.org/.}}

\section{Introduction}\label{sec:intro}

Over the last few years, biology has experienced a revolution as the
result of the introduction of new DNA sequencing technology, 
known as Next-Generation Sequencing (NGS), that nowadays makes it possible to sequence the genomic DNA or RNA transcripts, 
or transcriptome, in a few days instead of years, at a very low cost. 
These recent high-throughput sequencers produce data at unprecedented rates and scale, with associated sequencing costs 
in continuous decrease. In particular, RNA sequencing (RNA-seq) technology~\cite{rna_seq_tecnology} has arisen 
as a crucial analysis tool for biological and clinical research, as it can help to determine and quantify the expression 
of genes, the RNA transcripts, that are activated or repressed as the
result of different diseases or phenotypes, 
therefore providing an unbiased profile of a transcriptome that helps to understand the etiology of a disease. 
In consequence, RNA-seq is increasingly replacing conventional expression microarrays
in most practical scenarios~\cite{rna_seq_tecnology}.

Current NGS technology can sequence short DNA or RNA fragments, of length usually between 50 and 400 nucleotides (nts), 
though new sequencers with longer fragment sizes are being developed. Primary data produced by NGS sequencers consists of 
hundreds of millions or even billions of short DNA or RNA fragments which are called reads. The first step in NGS data processing in 
many comparative genomics experiments, including RNA-seq or genome resequencing~\cite{alignment_short_dna}, involves mapping the reads onto 
a reference genome, in order to locate the genomic coordinates where these fragments come from. This step constitutes an
extremely expensive process from the computational point of view. Furthermore, sensitivity is also a serious concern 
at this point~\cite{survey_sequence},
given that natural variations or error sequencing may occur, yielding frequent mismatches between reads and the reference genome, which increase
the computational complexity of the procedure. 

The mapping process is particularly more difficult for RNA-seq, as the genes 
in eukariotes may be split into small regions, called exons, that are separated by intron zones composed of thousands of nucleotides. 
Once the exons are transcribed to RNA, they are brought together to form the transcripts in a splicing process. Thus, 
when mapping reads from RNA transcripts onto a reference genome, it must be taken into account that these reads may contain a splice junction and, 
therefore, involve different exons, so that in practice they may lie thousands of nucleotides apart. This situation is referred to as a gapped alignment.

Recently, a variety of programs based on the Burrows-Wheeler Transform (BWT)~\cite{bwt} have been developed with the goal of accelerating 
the mapping process~\cite{soap2,bwt_long_read}. BWT is a popular pruning technique that has been successfully applied to accelerate 
ungapped alignment in genome index-based searches. A strategy that combines index-based read mapping with splice junction 
detection has been implemented in TopHat~\cite{tophat_sp}, a software package that is extensively used 
for the analysis of RNA-seq experiments. 
TopHat internally invokes Bowtie~\cite{alignment_short_dna}, a program for read mapping but with no support for 
gapped alignment. To tackle this in TopHat, 
read mappings that lie in close genomic locations are combined to reconstruct putative exon junctions, where unmapped reads are tried again. 
However, this approach presents performance and sensitivity problems, rooted in the {\em ad hoc} junction detection algorithm and the read mapping.
The underlying reason is basically that Bowtie only permits a small number of mismatches, 
which are insufficient for the current dimension of reads, therefore failing to map a 
large number of reads. Although Bowtie~2~\cite{bowtie2} significantly accelerates the mapping step, 
the strategy for junction detection remains unchanged in TopHat.

In this paper we contribute an innovative strategy that combines an
efficient algorithm for junction detection and a mapping algorithm
with notably improved sensitivity that correctly aligns reads with a
high rate of mutations (errors), insertions or deletions (EIDs).  This
strategy, implemented as a parallel pipeline in a program called
\rnaseq{}, shows excellent sensitivity and remarkable parallel
performance for both short and long RNA-seq reads, presenting runtimes
that linearly depend on the number and the length of reads.  In
\rnaseq{}, reads are aligned using a combination of mapping with BWT
and local alignment with the Smith-Waterman algorithm (SWA)~\cite{sw}.
As BWT is faster but less precise than SWA, we employ the former in
the early stages of the process, to obtain a rapid mapping of a large
number of reads (those which contain few EIDs); on the other hand, SWA
is applied in the final stages, to reliably map conflictive reads.
Furthermore, in our approach splice junctions are detected by dividing
unmapped reads into multiple seeds, which are next mapped using BWT
at distances compatible with the length of an intron. The potential
mapping regions detected using these seeds are then identified and
brought together to perform SWA alignment.

The rest of the paper is structured as follows.
In section~\ref{sec:pipeline} we describe the pipelined organization of \rnaseq{}
and the operation of the different stages.
In section~\ref{sec:parallel} we review the parallelization of the pipeline 
using OpenMP, a standard for portable shared memory parallel programming~\cite{openmp}.
A detailed experimentation is reported next, in section~\ref{sec:experiments},
exploring the performance of the stages and the full pipeline from the perspective 
of both parallelism and scalability on a server equipped with 64 AMD cores;
moreover, in that section we also include a comparison of \rnaseq{} against TopHat~2 on a platform with 12 Intel cores.
We finally close the paper with a discussion of conclusions and future work in section~\ref{sec:conclusions}.


\section{The Mapping Pipeline}\label{sec:pipeline}

\begin{figure*}
  \begin{center}
    \includegraphics[width=.7\linewidth]{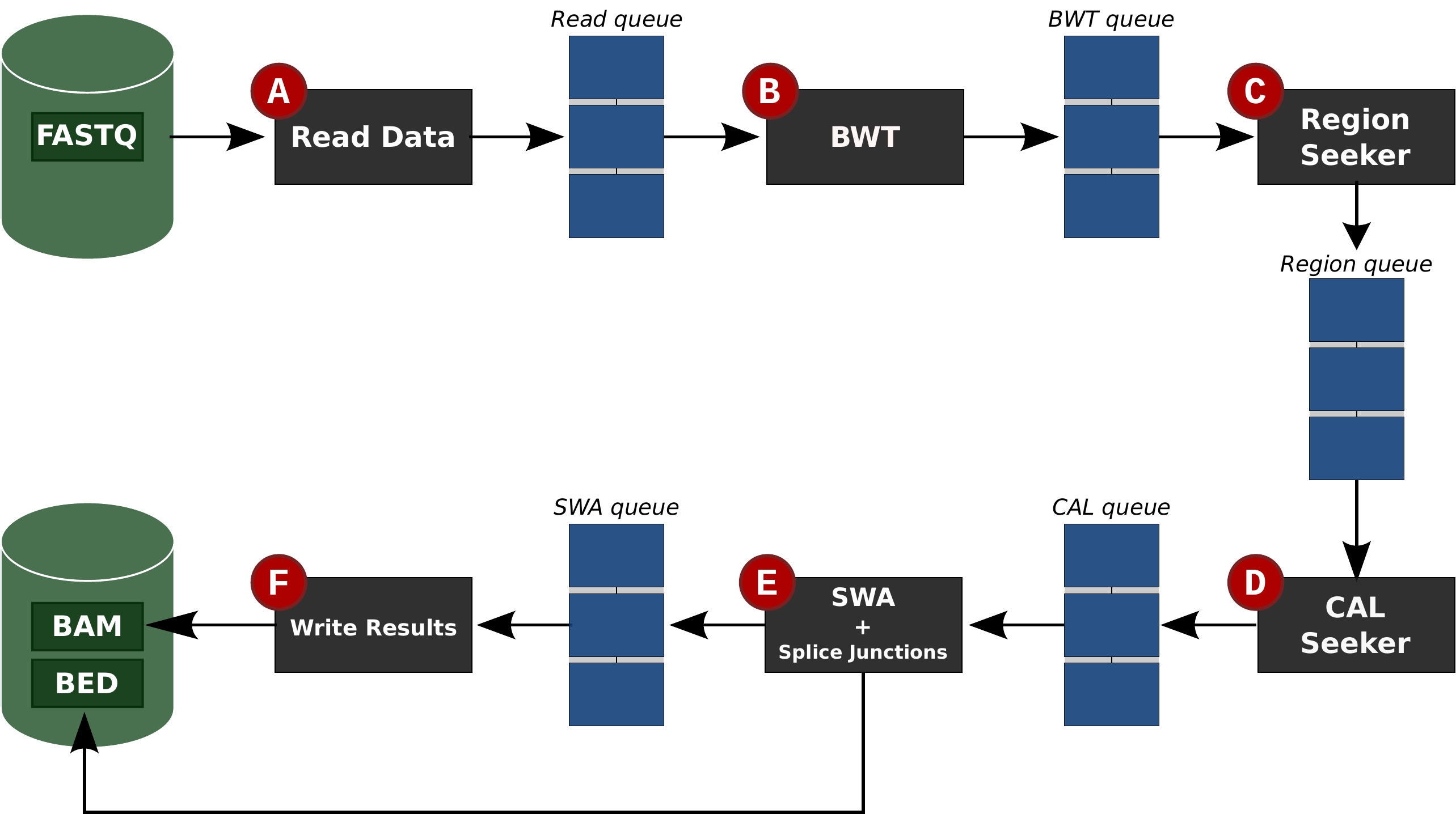}
    \caption{Pipelined organization of the RNA mapper.}
    \label{fig:pipeline_diagram}
  \end{center}
\end{figure*}

The \rnaseq{} pipeline maps RNA sequences into the reference genome,
with the mapping process being divided into the stages \emph{A--F}
illustrated in Figure~\ref{fig:pipeline_diagram}. The interaction between two
consecutive stages follows the well-known producer-consumer relation,
and is synchronized through a shared data structure, where the
producer inserts work for the consumer to process. In the following
paragraphs we discuss the tasks performed in each one of these stages together
with relevant algorithmic details.
%
We note that three of the pipeline stages ($B$, $C$ and
$E$) heavily rely on BWT and SWA.  

\subsection{Stage $A$: Read Data}
Initially the data corresponding to the RNA reads are stored in a
disk file following the standard FASTQ format~\cite{fastq}.  Due to
the large number of reads involved per experiment (typically, dozens
of millions) and the information that needs to be maintained per read,
this file is quite big, and in general exceeds the capacity of the
main memory. (Although the file sizes
can vary much depending on the case study, in our experiments we often
had to deal with files of 40--50 GBytes, which obviously will not
fit in the memory of most of todays' desktop servers.)

Therefore the principal task of this stage consists in retrieving data
from the disk in blocks, hereafter referred to as \emph{batches of
  reads}, with about 200 reads per batch, that are then stored in the
\emph{Read queue} for latter consumption by the subsequent stage.  The
batches of reads are kept in main memory in an array list, which
accommodates fast serial and indexed access. Among other information,
this array records the header, sequence, size, and quality of each
read.

\subsection{Stage $B$: BWT}
This stage performs a fast mapping of reads to the genome, using
our own implementation of the BWT.
The procedure extracts a batch from the read queue, and then applies
the BWT-based algorithm, allowing up to 1~EID per read.

If the read is successfully mapped, this stage creates an {\em
  alignment record} for each mapping that identifies the chromosome,
among other information, with the initial and final positions of the
read within the chromosome, and the strand.  Otherwise, the
information that identifies the unmapped read is stored in a
\emph{target array}.  The alignment records and target array conform a
single data structure with the batch of reads, that is passed to the
next stage once all the reads of the batch have been processed, via
the \emph{BWT queue} (which contain both mapped and unmapped reads).

\fixedwidthfigure{A read split into multiple overlapping seeds.}{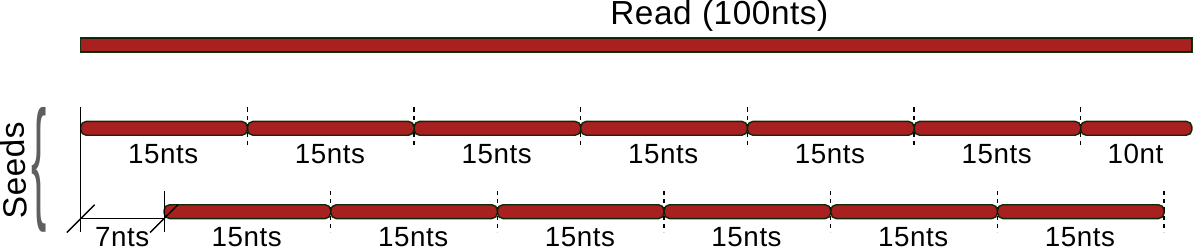}

\subsection{Stage $C$: Region Seeker}
Given a batch in the BWT queue, this stage only processes the unmapped
reads, leaving untouched those reads that were already mapped in
the previous stage. Each unmapped read of the batch is split in this
stage into a number of overlapping ``fragments'', hereafter referred
to as \emph{seeds}, tentatively of 15 nts each. For
example, given a read comprising 100 nts,
15~overlapping seeds can be
obtained: 8~seeds by starting the splitting at the first nucleotide of
the read, and 7~seeds more by starting the splitting at the eight
nucleotide of the read (see Figure~\ref{fig:seeds_from_read}). 

Next, the BWT-based algorithm is employed again to map each one of these
seeds into the reference genome, though this time no EIDs are
permitted.  The rationale of this process is the following. The most
likely reason a given read was not mapped in stage~\emph{B}
is that it contained more than one~EID. By dividing the read into
shorter seeds, we expect that all the EIDs of the read are
concentrated in only a few of these seeds. Therefore, the majority of
the seeds will be successfully mapped into the reference genome with no
EIDs.

Note that the seeds are quite small, leading to a new problem as it is
now very likely that each seed will be mapped to more than one place
in the reference genome.
The result of mapping these seeds is therefore a large collection of
what we call \emph{regions}, which identify all the places in the
reference genome where these seeds were successfully mapped.

Once the full batch of reads is processed, the results are stored as
part of the same data structure and passed to the next stage via the
\emph{Region queue}.

\subsection{Stage $D$: CAL Seeker}
This stage also processes the input information by batches, skipping the reads
that were already mapped in stage \emph{B}. For each unmapped read in
the batch, this stage uses the regions produced by stage \emph{C} to
obtain a list of \emph{candidate alignment locations} (CALs), which
define potential mappings of that read.

Let us briefly elaborate on this. As stated previously, for each read, the previous stage will have likely
detected a considerable number of regions due to the multiple matchings of the corresponding individual
seeds. Nevertheless, we expect than only those regions that are
related to a correct mapping of the read lie close together. We
consider these areas as potential mappings of some part of the
read, and therefore we identify them as CALs.

In particular, to obtain the CALs for a given read, this stage first merges the
regions of one read that are less than a certain number of nucleotides
apart; and then classifies each merged region as a CAL only if its
number of nucleotides is larger than a given
threshold. Figure~\ref{fig:cals} shows two CALs, $CAL_0$ and
$CAL_1$, that have been identified from the regions obtained by mapping
the seeds of a given read. In this figure, the gaps in the regions
represent the places where some of the seeds could not be
properly matched to the reference genome.

\fixedscalefigure{Identification of CALs from regions.}{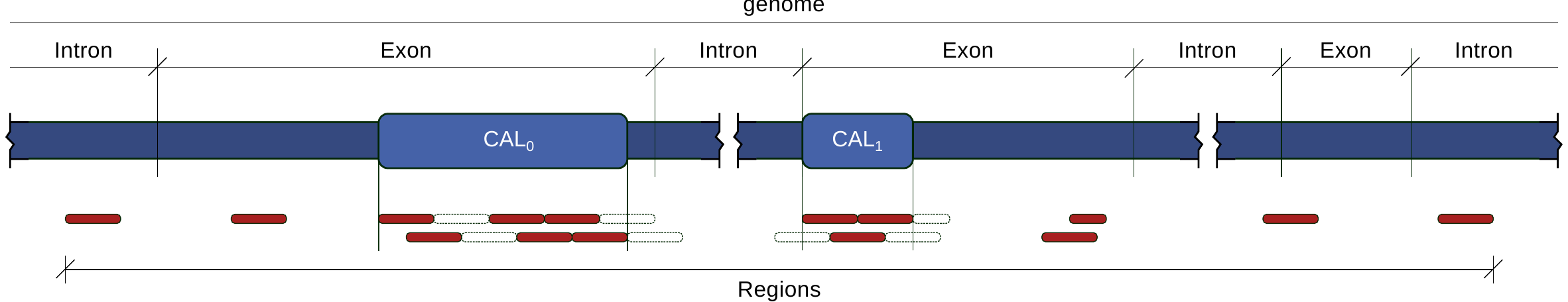}

At the end of this stage, the CALs are recorded in a data structure
which is passed to the next stage via the \emph{CAL queue}.
Furthermore, to save space, the information about the regions is
deleted at this point, as it will not be used in subsequent stages.

\subsection{Stage $E$: SWA + Splice Junctions}
This stage performs five consecutive tasks:
\begin{inparaenum}[i)]
\item CALs are extended with additional information from
  the reference genome;
\item extended CALs which are close to each other in the reference
  genome are joined;
\item the current read is mapped onto all extended CALs (joined or not),
  using our implementation of SWA;
\item successful mappings are recorded into a data structure and
  splice junctions are detected; and, finally,
\item splice junctions are written to disk.
\end{inparaenum}
The following paragraphs describe these five tasks in more detail.

The first task extends the CALs in order to include fragments from the reference genome that 
help to completely align a read. The CALs of a read are
extended to both left and right by a fixed number of nucleotides (30~nts by
default), though this could lead to intron fragments being included
in the extended CALs. This process is graphically illustrated in the
upper part of Figure~\ref{fig:extend_cals}.

The second task connects extended CALs which are close to each
other in the reference genome.  All the extended CALs whose distance
to each other is shorter than the longest known intron are merged,
generating a \emph{joined extended CAL}; see
Figure~\ref{fig:extend_cals}. 
Note that joined extended CALs involving more than two CALs are also possible.

\fixedscalefigure{A joined extended CAL obtained by joining two extended CALs
  that are close to each other.}{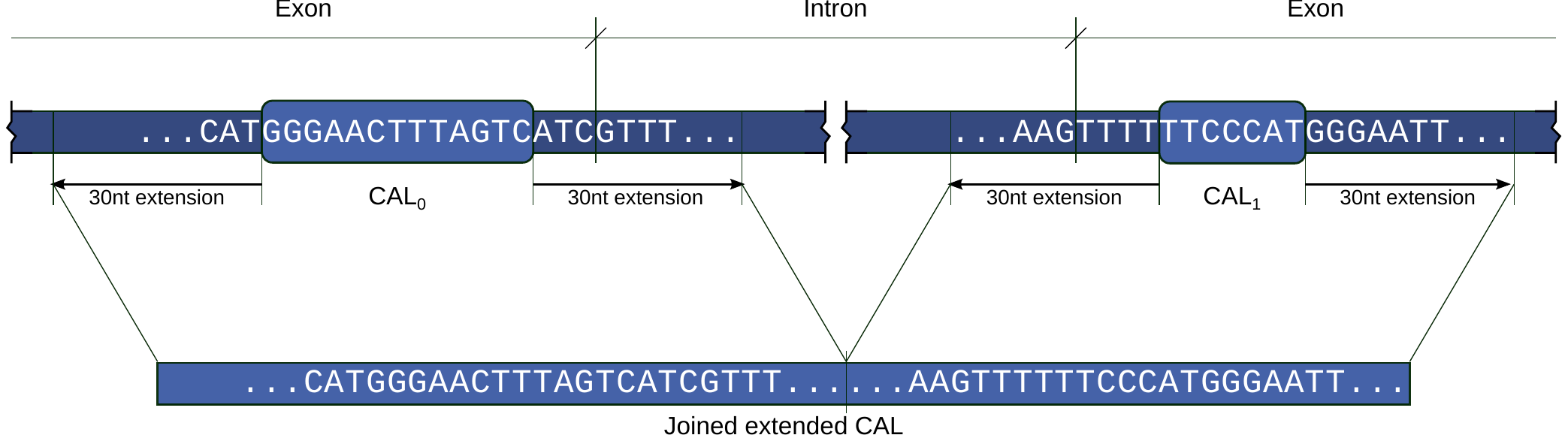}

After these two tasks, we will have identified a collection of extended CALs and
joined extended CALs, which are potential mapping candidates for the given
read. The following task maps the read onto all of these,
employing our own implementation of SWA. This task returns the alignment
between the read and each extended CAL, the statistics of the
alignment (e.g., quality score), and the initial
position of the alignment.

The fourth task is only performed on those SWA alignments
with alignment scores above a given threshold, that indicates a
successful mapping. In those cases, this task creates an alignment
record that identifies the chromosome, the initial and final positions
of the read within the chromosome, the strand, etc.  These alignment
records are passed to the next stage via the \emph{SWA queue}.

This fourth task also searches for splice junctions. For this purpose,
a large number of consecutive deletions in the alignment returned by
SWA is used as an indication of a potential splice junction; see
Figure~\ref{fig:splicing}. We expect that these deletions are due to
the intron fragments that were included in the joined reference when
the CALs were extended. Therefore, to assess if a large gap is an
actual splice junction, the start and end marks of an intron are
searched for the specific nucleotide sequences ``GT-AG'' and
``CT-AC''.

\fixedscalefigure{Splice junction detection.}{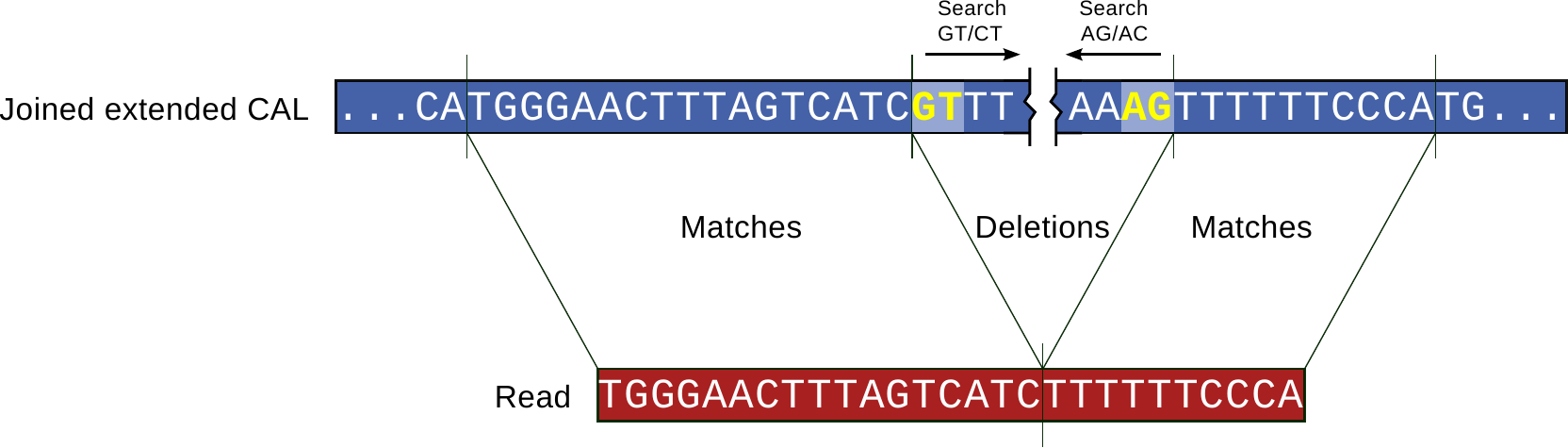}

All the splice junctions and the number of times each of them
has been detected are written to disk when the batches are
processed. Therefore, this information must be maintained and updated
during the whole process (i.e., not only for the current batch). To
reduce the memory space needed to hold the information on detected splice junctions,
as well as to rapidly find whether the last
detected splice junction was previously detected, a
self-balancing binary search tree (an AVL tree) data structure is used
to hold this information.

The last task of this stage simply writes the detected splice
junctions to an output file on disk, with the BED
format~\cite{bed}. As stated before, this information is written
to disk only when the last batch has been processed.

\subsection{Stage $F$: Write Results}
This stage completes the processing of a read batch, writing to disk
whether each read was mapped or not, following the BAM
format~\cite{bam_sam_api}. The output file contains, among other
information, the read id (the first component of the header), the
sequence, and its quality. In addition, whether the current read was
mapped, the chromosome, and the initial position are also written
to the output file.


\section{Leveraging Inter-stage and Intra-stage Concurrency}\label{sec:parallel}

\begin{figure*}[htbp!]
  \centering
  \subfloat{\includegraphics[width=.9\linewidth]{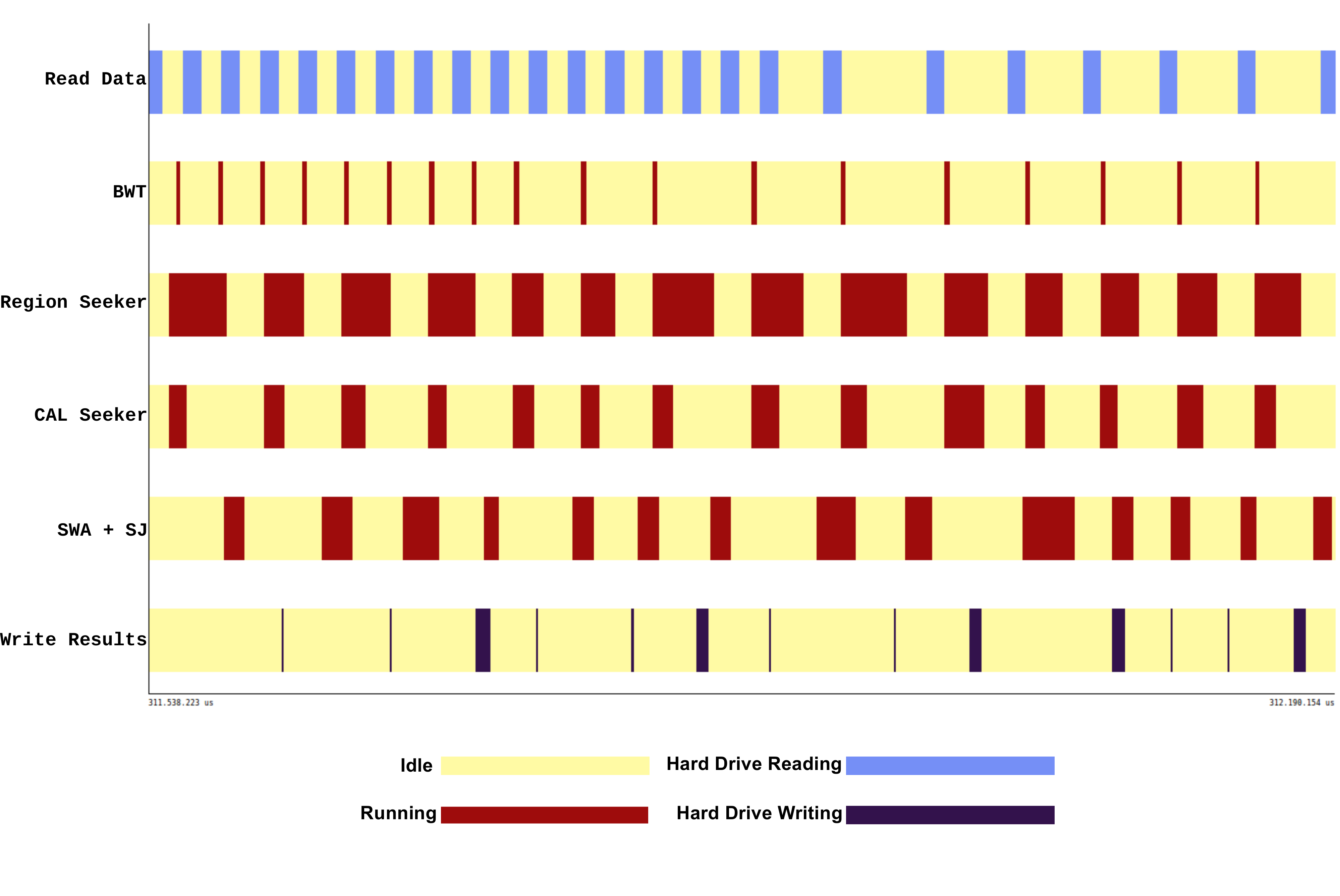}}
  \\
  \subfloat{\includegraphics[width=.9\linewidth]{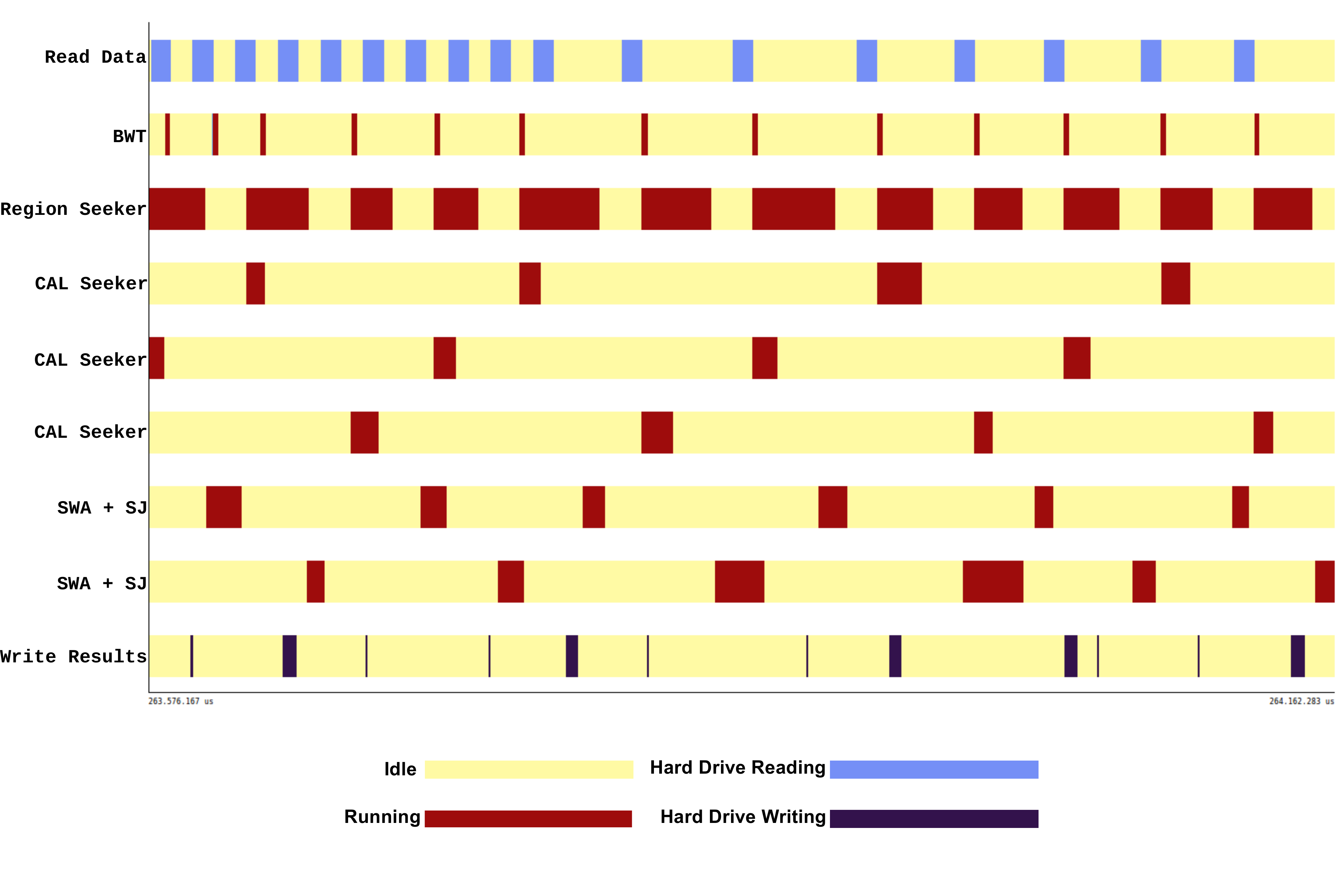}}
  \caption{Traces of a pipelined execution of the mapper exploiting
    inter-stage concurrency (top) and both inter- and intra-stage
    concurrency (bottom). In the first case, a single thread is
    employed for each one of the pipeline stages: {\tt Read data},
    {\tt BWT}, {\tt Region seeker}, {\tt CAL seeker}, {\tt SWA+SJ},
    and {\tt Write Results}. The execution thus yields an interleaved
    execution with, e.g., I/O from/to disk proceeding in parallel with
    the other stages. The second case goes one step further in the
    exploitation of concurrency, employing, in this particular
    example, 3 threads for the {\tt CAL seeker} and 2 for the {\tt
      SWA+SJ}. The net result is that now several batches can be
    processed in parallel by different threads, even in the same
    stage.}
  \label{fig:pipeline_trace}
\end{figure*}

The pipelined organization of our \rnaseq{} mapper naturally accommodates two types of algorithmic concurrency,
both targeting the hardware parallelism of current multicore processors.
On one side, the mapper is divided into a number of independent stages, connected
via queues that act as data buffers and synchronize the relative processing speeds (throughput) of consecutive stages.
Provided each one of these stages is run on a separate thread and a different (CPU) core, 
the result is an overlapped execution of the stages ---see the execution trace in Figure~\ref{fig:pipeline_trace} (top)---, 
much like the exploitation of
instruction-level parallelism that occurs in a pipelined processor~\cite{HenP12}.
The actual benefits that this pipelined organization yields, in terms of reduced execution
time, are experimentally analyzed in the next section. 

On the other side, in case there exists a sufficient number of computational resources (basically,
cores), nothing prevents our scheme from also leveraging the concurrency intrinsic to each one of the stages, 
so that more than one thread can operate per stage,
mimicking the operation of a pipelined superscalar processor~\cite{HenP12}; see the execution
trace in Figure~\ref{fig:pipeline_trace} (bottom).
The purpose of exploiting intra-stage concurrency is to both accelerate and  equilibrate the processing speeds of the
different stages, as the throughput of the full pipeline is dictated by the performance of the slowest stage.  

We next discuss the general parallel pipelined framework as well as the specific approaches that were adopted in our scheme 
to exploit concurrency within stages $B$--$E$. 
Stages $A$ and $F$ both perform I/O from/to the disk and, thus, there is not much to be gained from a parallel execution on a desktop
server like that targeted by our software.
Although we will show that the parallelization of the framework and stages using OpenMP~\cite{openmp} is quite natural, we note that
this is only possible because of the careful organization of the complete \rnaseq{} mapping pipeline, 
specifically, the design of clean procedural interfaces and specialized shared data structures.\\

\subsection{General parallel framework} 

Listing~\ref{lst:framework} illustrates the parallelization of the pipelined \rnaseq{} mapper using OpenMP.
The organization of the algorithm into 6 independent stages (see Figure~\ref{fig:pipeline_diagram}), 
connected via synchronization buffers, allows us to
 employ the OpenMP {\bf sections} construct, with one thread dedicated to run each one of the stages, in principle yielding an overlapped
(and, therefore, parallel) execution; see Figure~\ref{fig:pipeline_trace} (top).
For simplicity, we do not include certain details specific to the pipeline operation.
For example, condition variables are employed in the actual implementation to synchronize consecutive stages,
forcing the producer/consumer to stop when the shared queue is full/empty.
Also, the operation of one particular stage is terminated when its input queue is empty and the previous stage indicates
that it will produce no further new items (batches).\\

\begin{lstlisting}[language=C,caption=Simplified version of pipelined framework
                   parallelized using OpenMP.,label=lst:framework,
                   morekeywords={pragma,omp,parallel,shared,private,sections,section,num_threads},float=*]
#define num_stages    6
int AB_pending_work = TRUE,
    BC_pending_work = TRUE,
    CE_pending_work = TRUE,
    EF_pending_work = TRUE;

#pragma omp parallel sections num_threads( num_stages ) {
  #pragma omp section {
    A_Read_Data( FASTQ_file, Read_queue );          // Read batches from FASTQ 
  }
  #pragma omp section {
    B_BWT( Read_queue, BWT_queue );                 // Apply the BWT
  }
  #pragma omp section {
    C_Region_Seeker( BWT_queue, Region_queue );     // Apply Region seeker
  }
  #pragma omp section {
    D_CAL_Seeker( Region_queue, CAL_queue );        // Apply CAL seeker
  }
  #pragma omp section {
    E_SWA_Splice_Junctions( CAL_queue, SWA_queue ); // Apply SWA and 
                                                    // detect splice jcts
    E_Write_Splice_Junctions( BED_file );           // Write results onto BAM 
  }
  #pragma omp section {
    F_Write_Data( SWA_queue, BAM_file );            // Write results onto BAM
  }
}
\end{lstlisting}

\subsection{Stages $A$ and $F$: Read Data and Write Results}

These stages stream the data between the disk and the pipeline in batches of reads.
They both proceed sequentially, as no attempt is made within each one of these stages to read/write
multiple batches from/to disk simultaneously. 

Listing~\ref{lst:stagesAF} shows a simplified version of the process performed within these stages.
Note how the procedure for stage~$A$ moves batches from the input file {\tt FASTQ\_file} to the 
``output'' {\tt Read\_queue}, 
till the end of the file is detected.
At that point, this stage notifies the following one that it will not produce new batches by setting 
{\tt AB\_pending\_work} to {\tt FALSE}.

The code for stage~$F$ performs the opposite process, moving data from the {\tt SWA\_queue} to the 
file containing the mapping information, {\tt BAM\_file}, till there is no more {\tt EF\_pending\_work} (notified by the previous
stage) and the ``input'' {\tt SWA\_queue} is empty.\\

\begin{lstlisting}[language=C,caption=Simplified version of serial reads and writes of batches within stages~$A$ and $F$
                   respectively
                   using OpenMP.,label=lst:stagesAF,
                   morekeywords={pragma,omp,parallel,shared,private,sections,section,num_threads},float=*]
void A_Read_Data( FASTQ_file_t FASTQ_file, Read_queue_t Read_queue ) {
  while ( !end_file( FASTQ_file ) ) {
    A_Read_Batch(  FASTQ_file, Batch ); // Read a single batch from 
    A_Write_Batch( Read_queue, Batch ); // FASTQ file onto Read_queue
  }
  AB_pending_work = FALSE;
}

void F_Write_Data( SWA_queue_t SWA_queue,  BAM_file_t BAM_file ) {
  while ( ( EF_pending_work ) OR ( !empty_queue( SWA_queue ) ) ) {
    F_Read_Batch(  SWA_queue,  Batch ); // Write a single batch from SWA_queue
    F_Write_Batch( BAM_file,   Batch ); // onto BAM file
  }
}
\end{lstlisting}

\subsection{Stages $B$ and $C$: Processing reads within a batch in parallel} 

Given a batch of reads, 
these two stages exploit that any two reads can be processed completely independently.
Thus, in stage~$B$ we dedicate {\tt nt\_BWT} OpenMP threads to concurrently apply the BWT-based algorithm 
(with up to one EID) 
to process the reads of a batch (nested parallelism), with a straight-forward implementation that simply adds 
a loop-parallel OpenMP directive to the serial code; see the sample
code in Listing~\ref{lst:stagesBC}.
Termination is detected using
the variable {\tt AB\_pending\_work} (set by the previous stage) and the test on the contents of the {\tt Read\_queue}.
Furthermore, detection of termination is notified to the subsequent stage by setting 
{\tt BC\_pending\_work} to {\tt FALSE} upon completion.

\begin{lstlisting}[language=C,caption=Simplified version of parallel processing of reads within stages~$B$ and~$C$
                   using OpenMP.,label=lst:stagesBC,
                   morekeywords={pragma,omp,parallel,shared,private,sections,section,num_threads},float=*]
#define nt_BWT           ... // User-defined parameter
#define nt_Region_Seeker ... // User-defined parameter

void B_BWT( Read_queue_t Read_queue, BWT_queue_t BWT_queue ) {
  while ( ( AB_pending_work ) OR ( !empty_queue( Read_queue ) ) ) {
    B_BWT_Read( Read_queue, Batch );
    #pragma omp parallel for num_threads( nt_BWT ) {
      for ( i = 0; i < Batch.n_reads; i++ )   // Loop over all reads 
        B_BWT_Single_Read( Batch.read[ i ] ); // Apply the BWT to the i-th read 
    }
    B_BWT_Write( BWT_queue, Batch );
  }
  BC_pending_work = FALSE;
}

void C_Region_Seeker( BWT_queue_t BWT_queue, Region_queue_t Region_queue ) {
  while ( ( BC_pending_work ) OR ( !empty_queue( BWT_queue ) ) ) {
    C_Region_Read( BWT_queue, Batch );
    #pragma omp parallel for num_threads( nt_Region_Seeker ) {
      for ( i = 0; i < Batch.n_reads; i++ )      // Loop over all reads
        C_Region_Single_Read( Batch.read[ i ] ); // Apply Region seeker 
                                                 // to the i-th read
    }
    C_Region_Write( Region_queue, Batch );
  }
  CD_pending_work = FALSE;
}
\end{lstlisting}

Note that each batch usually consists of 200~reads and, therefore, given the moderate number of cores available
in current desktop servers, there exists a significant amount of concurrency per batch.
Also, no special care is needed during the concurrent access to the data structure 
containing the batch of reads, as each thread will only modify the registers corresponding to
a distinct read and, therefore, no race conditions are possible.

Similar comments apply to the search of regions performed in stage~$C$: a simple OpenMP-based implementation 
suffices to exploit nested parallelism; and coordination with previous and subsequent stage is performed in an analogous manner,
so that no race conditions can result from the concurrent accesses to the data structures.
However, note that now the degree of concurrency within
a batch of reads is in general more reduced, as this stage only processes unmapped reads from the previous stage.\\

\subsection{Stages $D$ and $E$: Processing batches of reads in parallel} 

These stages exploit the independence among the information
contained in the batches to process them concurrently (nested parallelism), with 
{\tt nt\_CAL\_Seeker} and
{\tt nt\_SWA\_Splice\_Junctions} threads (for stages $D$ and $E$, respectively), 
using the {\tt parallel} OpenMP directive.
The degree of parallelism available at each one of these stages depends on the throughput of the previous stage.
For example, provided stage~$C$ processes batches at sufficient speed, inserting these results onto the
{\tt Region\_queue}, stage~$D$ will receive a constant flux of inputs, enough to feed the computational resources (threads) dedicated to
search the CALs. Note again that the autonomy of the batches ensure that no special mechanism is necessary to avoid
race conditions.

\begin{lstlisting}[language=C,caption=Simplified version of parallel processing of batches within stages~$D$ and~$E$
                   using OpenMP.,label=lst:stagesDE,
                   morekeywords={pragma,omp,parallel,shared,private,sections,section,num_threads},float=*]
#define nt_CAL_Seeker           ... // User-defined parameter
#define nt_SWA_Splice_Junctions ... // User-defined parameter

void D_CAL_Seeker( Region_queue_t Region_queue, CAL_queue_t CAL_queue ) {
  #pragma omp parallel private( Batch ) num_threads( nt_CAL_Seeker ) {
    while ( ( CD_pending_work ) OR ( !empty_queue( Region_queue ) ) ) {
      D_CAL_Seeker_Read(  Region_queue, Batch );
      D_CAL_Seeker(                     Batch ); // Apply CAL seeker to all reads 
      D_CAL_Seeker_Write( CAL_queue,    Batch ); // within the batch
     }
  }
  DE_pending_work = FALSE;
}

void E_SWA_Splice_Junctions( CAL_queue_t CAL_queue, SWA_queue_t SWA_queue ) {
  #pragma omp parallel private( Batch ) num_threads( nt_SWA_Splice_Junctions ) {
    while ( ( DE_pending_work ) OR ( !empty_queue( CAL_queue ) ) ) {
      E_SWA_Splice_Junctions_Read(  CAL_queue, Batch );
      E_SWA_Splice_Junctions(                  Batch ); // Apply SWA/detect splice
      E_SWA_Splice_Junctions_Write( SWA_queue, Batch ); // jcts to/in all reads 
                                                        // within the batch
    }
  }
  EF_pending_work = FALSE;
}
\end{lstlisting}



\section{Experimental Results}\label{sec:experiments}

The experiments reported in the next three  subsections were performed
on a platform equipped with four AMD Opteron 6276 processors at 2.3
GHz, with 16 cores each (64 cores in total), and 64 Gbytes of RAM.
This is a ccNUMA (cache-coherent non-uniform memory access)
architecture where each processor has its own memory controller and
local memory, and exchanges information with other processors via the
HyperTransport interconnect.

In the following subsections we only study the parallelism of stages $B$--$E$. 
Due to its (intra-stage) sequential nature, stages $A$ and $F$ are executed by a single thread
each. Independent experiments proved that the runtime of these two stages is 
negligible and, furthermore, their execution can be overlapped with the rest of the stages so that
they exert a minor impact on the performance of the full pipeline.

\subsection{Benchmarks}

The experimental evaluation of the aligner is performed
using simulated single-end datasets with 2 million reads of 100 and 250 nts.
The former length is representative of recent RNA-seq while, with the dropping costs of this technology, reads of longer
dimension (e.g., 250 nts) are  becoming mainstream.
Two scenarios were designed for variabilities ($\varepsilon$) of 0.1\% and 1\% of mismatches, with 30\% of them being insertions/deletions.
Hereafter, we will refer to the four benchmark test cases as 
{\tt T1} (100 nts, $\varepsilon=$0.1\%),
{\tt T2} (100 nts, $\varepsilon=$1\%),
{\tt T3} (250 nts, $\varepsilon=$0.1\%), and
{\tt T4} (250 nts, $\varepsilon=$1\%).
Simulations were carried out with the {\tt dwgsim} program, from the SAM tools~\cite{bam_sam_api},
setting the appropriate values for
{\tt n} (maximum number of Ns allowed in a given read),
{\tt r} (rate of mutations), and
{\tt c} (mean coverage across available positions).

\subsection{Intra-stage concurrency}

Our first study investigates the parallelism and scalability of the algorithms
employed in stages $B$--$E$.
Consider one of this stages, e.g., $D$.
In order to avoid interferences in this evaluation, we initially block (the threads in charge of executing) stage $D$ and
run all stages of the pipeline prior to it (i.e., $A$--$C$) to completion, so that the Region queue, between $C$ and $D$
(see Figure~\ref{fig:pipeline_diagram}),
is filled with the same data that the pipelined execution would produce for each test case.
We next block the stages after $D$ (i.e., $E$ and $F$) and let this one operate 
in isolation, with a variable number of threads executing it.
The results thus illustrate the strong scalability of the algorithm employed in the stage
in the ideal case when other stages are disabled and, therefore, no memory conflicts occur with these other stages.

\begin{table*}[th\!]
  \centering
  \caption{Execution time (in s.) of stages $B$--$E$ for the four test cases.}
  \label{tab:intrastage}
  \subfloat[Test case \texttt{T1}: 100~nts, $\varepsilon=$0.1\%.]{ %
    \begin{tabular}{rrrrr}
      \toprule
      &
      \multicolumn{4}{c}{Time per stage}
      \\
      Threads &
      \multicolumn{1}{c}{$B$} & \multicolumn{1}{c}{$C$} & 
      \multicolumn{1}{c}{$D$} & \multicolumn{1}{c}{$E$}
      \\
      \midrule
       1  & 38.58 & 562.80 & 219.79 & 318.02 \\
       2  & 23.20 & 254.14 & 110.53 & 161.19 \\
       4  & 16.00 & 143.42 &  57.56 &  84.30 \\
       6  & 11.89 & 101.87 &  40.45 &  56.12 \\
       8  & 11.31 &  81.52 &  31.15 &  43.07 \\
      10  & 10.14 &  69.34 &  26.42 &  35.46 \\
      12  &  9.62 &  64.69 &  22.52 &  30.83 \\
      14  &  9.26 &  58.65 &  19.72 &  27.43 \\
      16  &  9.70 &  60.78 &  18.84 &  25.87 \\
      18  & 10.01 &  69.00 &  16.82 &  22.62 \\
      20  & 10.04 &  57.25 &  15.95 &  22.05 \\
      \bottomrule
    \end{tabular}
  }
  \qquad{}
  \subfloat[Test case \texttt{T2}: 100~nts, $\varepsilon=$1\%.]{ %
    \begin{tabular}{rrrrr}
      \toprule
      &
      \multicolumn{4}{c}{Time per stage}
      \\
      Threads &
      \multicolumn{1}{c}{$B$} & \multicolumn{1}{c}{$C$} & 
      \multicolumn{1}{c}{$D$} & \multicolumn{1}{c}{$E$}
      \\
      \midrule
       1  & 32.42 & 615.99 & 243.33 & 361.26 \\
       2  & 22.20 & 347.88 & 126.31 & 179.54 \\
       4  & 14.43 & 168.50 &  65.24 &  89.35 \\
       6  & 10.37 & 114.26 &  45.71 &  61.46 \\
       8  & 10.08 &  90.33 &  35.36 &  48.95 \\
      10  &  9.21 &  78.58 &  29.49 &  38.88 \\
      12  &  8.88 &  72.06 &  25.06 &  33.19 \\
      14  &  8.31 &  64.36 &  22.84 &  29.49 \\
      16  &  8.71 &  63.68 &  20.74 &  28.17 \\
      18  &  8.89 &  68.23 &  18.95 &  25.09 \\
      20  &  9.00 &  59.55 &  18.00 &  23.81 \\
      \bottomrule
    \end{tabular}
  }
  \\
  \subfloat[Test case \texttt{T3}: 250~nts, $\varepsilon=$0.1\%.]{ %
    \begin{tabular}{rrrrr}
      \toprule
      &
      \multicolumn{4}{c}{Time per stage}
      \\
      Threads &
      \multicolumn{1}{c}{$B$} & \multicolumn{1}{c}{$C$} & 
      \multicolumn{1}{c}{$D$} & \multicolumn{1}{c}{$E$}
      \\
      \midrule
       1  & 37.07 & 1,545.78 & 730.98 & 2,198.11 \\
       2  & 23.05 & 1,039.40 & 367.63 & 1,098.15 \\
       4  & 17.11 &   529.66 & 192.09 &   550.30 \\
       6  & 13.34 &   391.12 & 129.66 &   368.46 \\
       8  & 12.06 &   336.35 & 100.11 &   280.41 \\
      10  & 11.44 &   290.42 &  84.55 &   235.23 \\
      12  & 11.75 &   249.79 &  71.23 &   202.54 \\
      14  & 10.50 &   230.41 &  62.66 &   177.94 \\
      16  & 11.49 &   206.89 &  57.87 &   158.29 \\
      18  & 11.43 &   210.77 &  53.39 &   149.09 \\
      20  & 11.69 &   198.17 &  48.82 &   138.32 \\
      \bottomrule
    \end{tabular}
  }
  \qquad{}
  \subfloat[Test case \texttt{T4}: 250~nts, $\varepsilon=$1\%.]{ %
    \begin{tabular}{rrrrr}
      \toprule
      &
      \multicolumn{4}{c}{Time per stage}
      \\
      Threads &
      \multicolumn{1}{c}{$B$} & \multicolumn{1}{c}{$C$} & 
      \multicolumn{1}{c}{$D$} & \multicolumn{1}{c}{$E$}
      \\
      \midrule
       1  & 36.67 & 1,872.21 & 711.24 & 2,149.98 \\
       2  & 37.89 & 1,058.06 & 366.99 & 1,074.69 \\
       4  & 24.00 &   544.68 & 190.43 &   541.59 \\
       6  & 16.20 &   380.14 & 130.88 &   367.99 \\
       8  & 13.68 &   301.08 &  99.37 &   277.76 \\
      10  & 12.77 &   265.31 &  82.56 &   233.05 \\
      12  & 12.28 &   238.11 &  71.67 &   197.50 \\
      14  & 11.60 &   226.34 &  62.75 &   171.05 \\
      16  & 12.16 &   221.27 &  58.05 &   160.21 \\
      18  & 12.71 &   222.73 &  53.00 &   146.68 \\
      20  & 13.01 &   223.02 &  49.01 &   135.84 \\
      \bottomrule
    \end{tabular}
  }
\end{table*}
 

\begin{figure*}[htbp!]
  \centering
  \subfloat{
    \includegraphics[width=.49\linewidth]{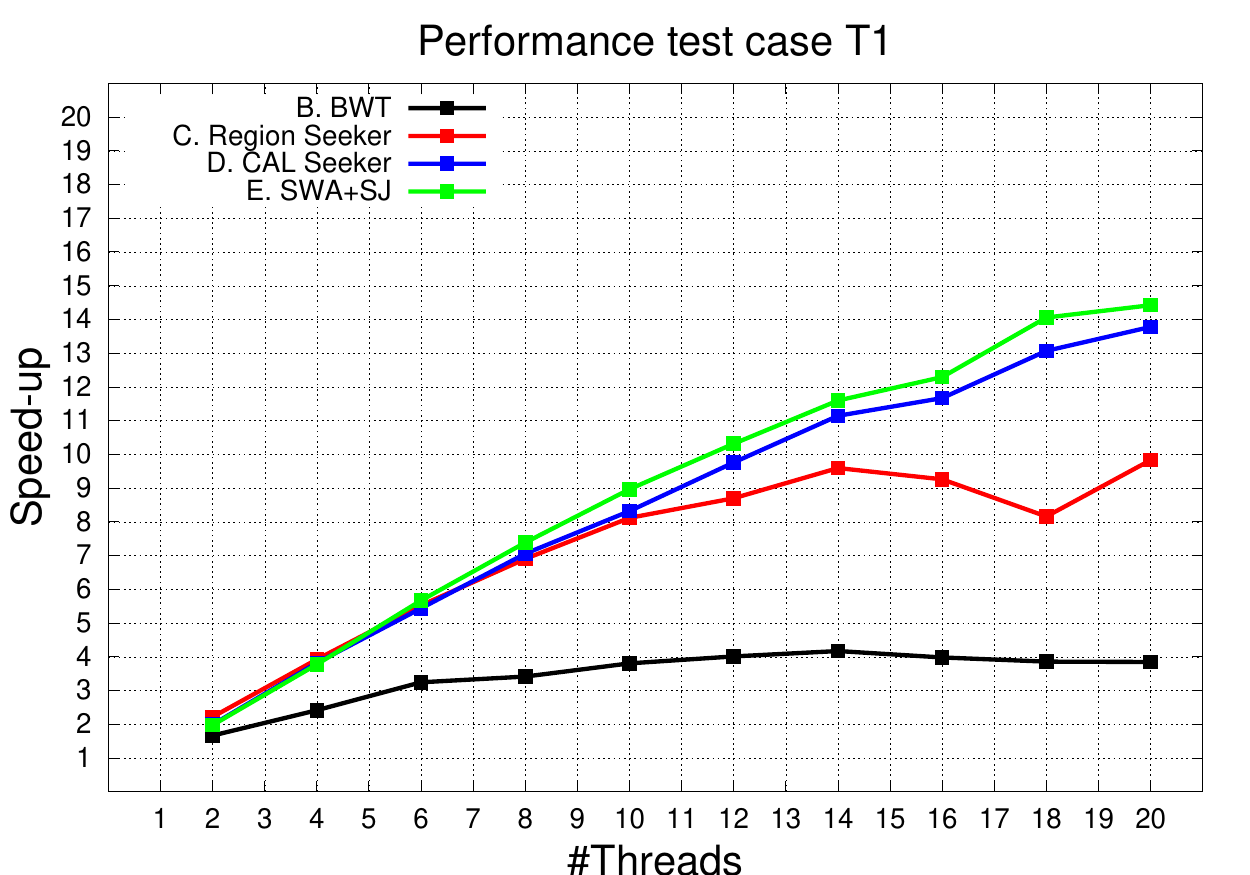}
  }
  \subfloat{
    \includegraphics[width=.49\linewidth]{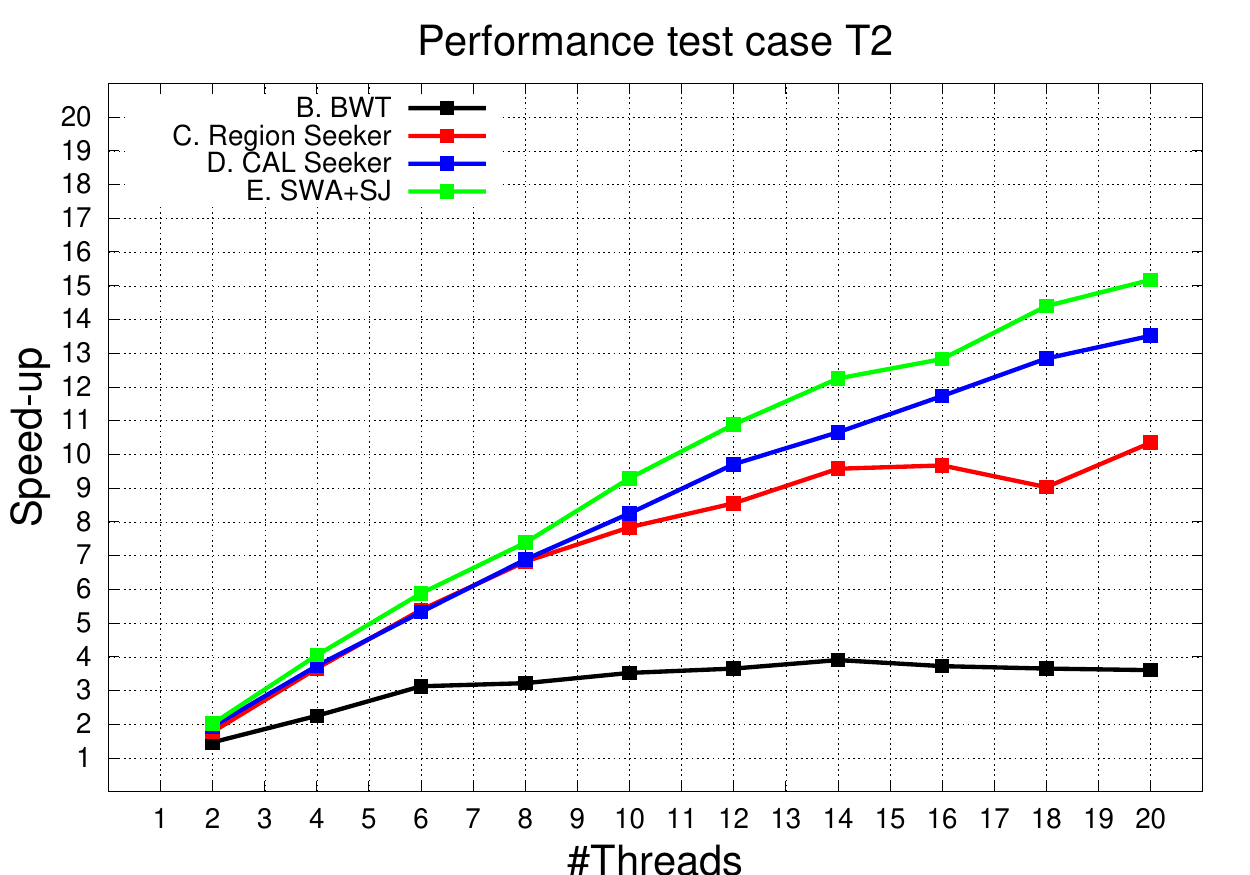}
  }
  \\
  \subfloat{
    \includegraphics[width=.49\linewidth]{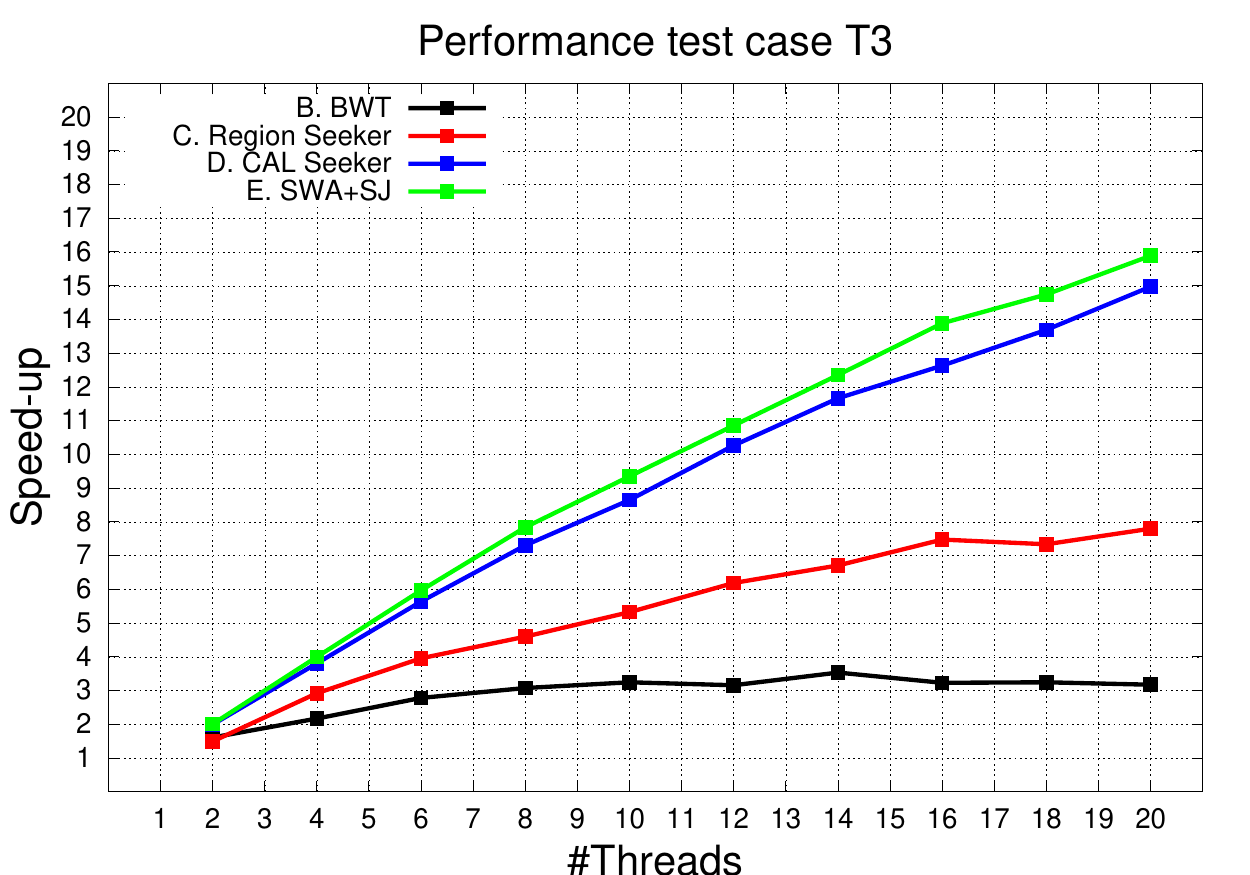}
  }
  \subfloat{
    \includegraphics[width=.49\linewidth]{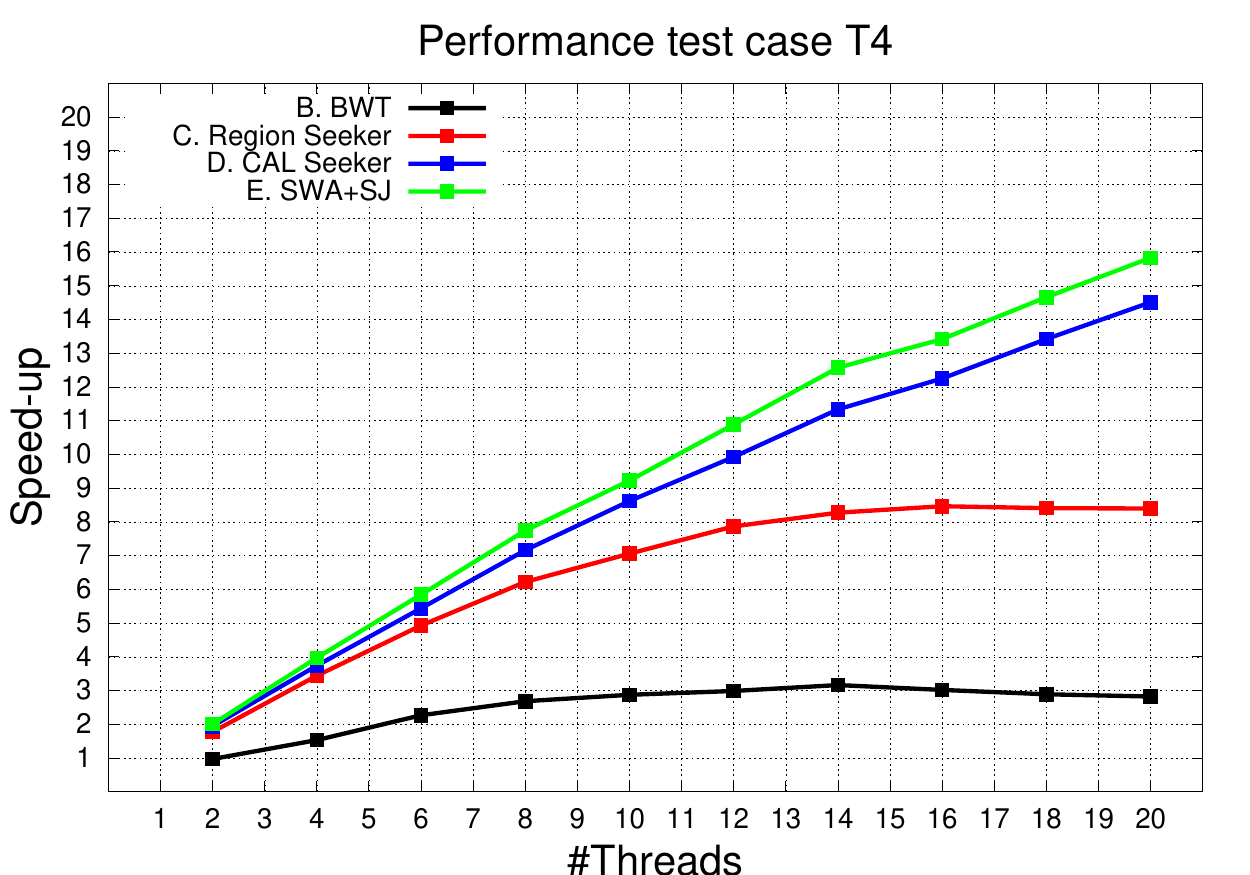}
  }
  \caption{Speed-ups of stages $B$--$E$ for the four test cases.}
  \label{fig:intrastage}
\end{figure*}

Table~\ref{tab:intrastage}
reports the execution time
of stages $B$--$E$ and test cases {\tt T1}--{\tt T4} when up to 20 threads are employed for the execution.
Figure~\ref{fig:intrastage} displays the corresponding speed-ups (obtained dividing the serial time, i.e., with
one thread, by the parallel time).
The first aspect to note from these results is 
the much higher cost of stage $C$ for the {\tt T1} and {\tt T2} cases, and the higher cost of
stage $E$ for the remaining two cases when a single thread is employed.

On the other hand, there is a similar parallel behaviour of the algorithms for all four 
benchmarks, with the BWT-based algorithm (stage $B$) showing the worst scalability, much below
those of the algorithms employed in the other three stages.
In principle, stage $B$ exhibits a perfect degree of parallelism: all reads within a batch are independent and,
therefore, they can be processed concurrently. To leverage this,
in our implementation of this stage each thread extracts a read from the input queue,
processes it, and inserts the result in the BWT queue.
However, the BWT presents an irregular access pattern to the data as well as a low computation-communication
(memory transfers) ratio. Furthermore, the implementation operates on a vector that does not fit into the cache memory,
rendering the BWT-based algorithm a memory-bounded process that is intrinsically limited
by the bandwidth of the interconnect to memory.
Stage~$C$ also employs the same algorithm, though this time it does so to map seeds
instead of complete reads. This partially explains the superior execution time of this stage as
each read that remained unmapped from the previous stage is now split into a collection of seeds to be
mapped. On the other hand, it seems that the fact that the seeds of a read will likely map in nearby areas
of the genome helps in reducing the amount of references to non-cached data, resulting in a higher speed-up for
this stage.
Finally, stages $D$ and $E$ show a good linear scalability that grows steadily with the number of threads.

\subsection{Inter- and Intra-stage concurrency}

Armed with the previous results,
we now study the performance of the full pipeline. 
Since the target platform has 64 cores, we could distribute (up to)
an analogous number of threads arbitrarily among the different
stages of the pipeline, reserving two cores/threads for the execution of stages $A$ and $F$ (I/O from/to disk,
respectively). However, the following two observations guided our choices.

First, the performance of a pipeline is limited by the slowest stage
so that it is convenient to assign the throughput of resources
carefully to equilibrate the execution time of the stages. In particular,
the execution times in Table~\ref{tab:intrastage} clearly ask for the application of a higher number of
threads to the execution of stages $C$--$E$ than to $B$. 
Depending on the test case, it is more convenient to dedicate more threads
to stage $C$ or $E$, though in any case these numbers should be superior than to those dedicated to $D$.
Note that these values are set through constants
{\tt nt\_BWT},
{\tt nt\_Region\_Seeker}\ldots{} %
in our codes.

The second observation is that the maximum number of threads that it is convenient to dedicate to certain 
stages is limited. For example, 
the analysis of the results Figure~\ref{fig:intrastage} reveals that assigning 
more than 4--6 threads for stage $B$ or, depending on the case, more than 12--16 threads for stage $C$, is useless.

\begin{figure}[htbp!]
\begin{center}
\includegraphics[width=\linewidth]{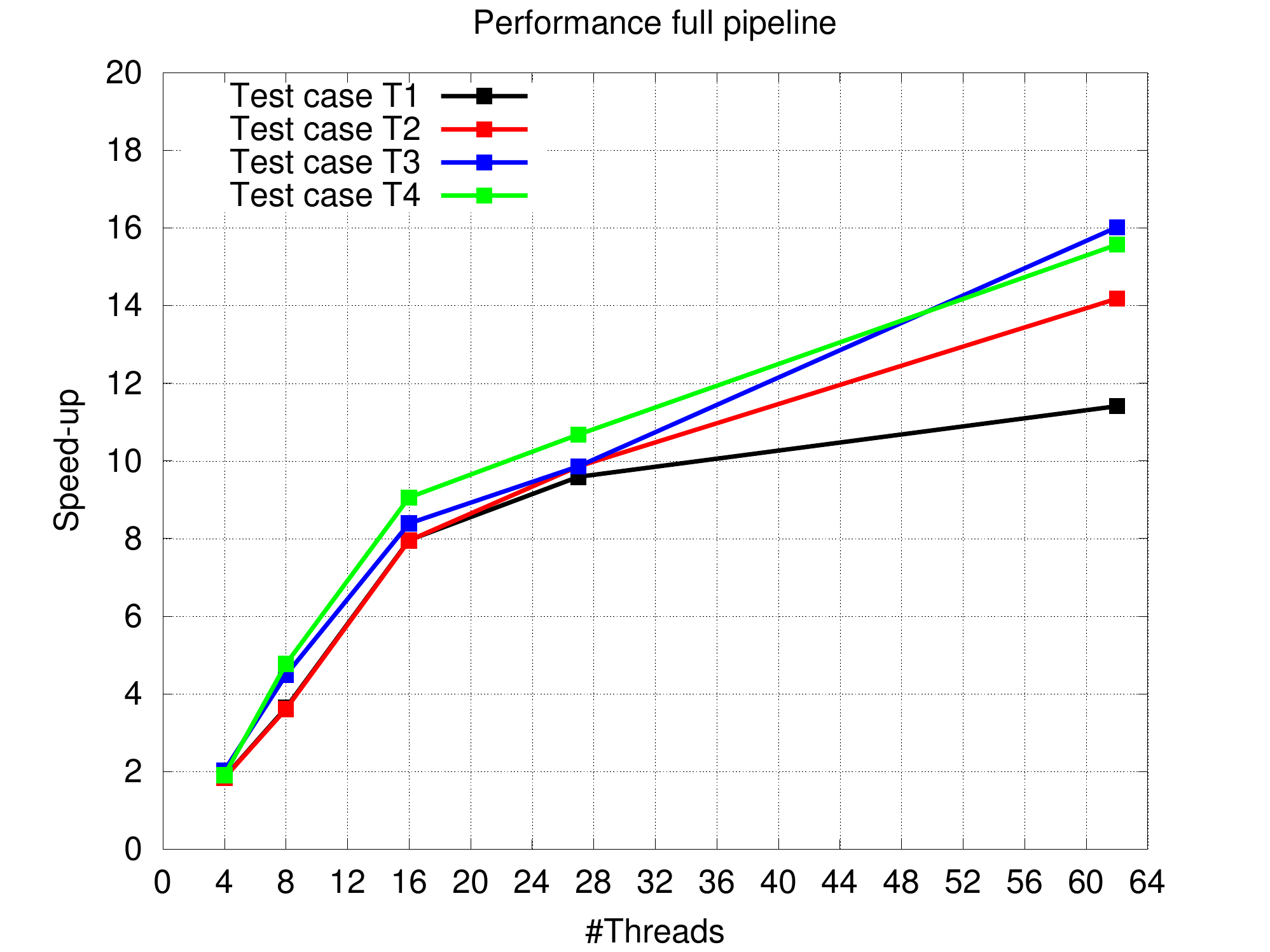}
\end{center}
\caption{Speed-up of the full pipeline for the four test cases.}
\label{fig:full_pipeline}
\end{figure}

Following these two principles, let us examine the performance of the parallel execution of the pipeline.
The complete results from this experiment are collected in Table~\ref{tab:full_pipeline} and summarized
in Figure~\ref{fig:full_pipeline}. The total time includes also I/O from/to disk so that the actual number of
threads employed in this experiment is that showed in the corresponding column plus the (two) I/O threads.
The execution times per stage only reflect the period of time when the thread of the 
corresponding stage is active, processing data, but excludes the period that the thread is idle, 
e.g., waiting because there is no work in the input queue.

Consider, e.g., the results for benchmark {\tt T1} in Table~\ref{tab:full_pipeline}.
The first row of this table, labelled as ``Sequential version'', reports the times measured in a sequential 
execution of the pipeline ---i.e., no overlapping--- (recall Table~\ref{tab:intrastage}), and the total time resulting from 
that. (For simplicity, the cost of I/O is not included in the serial execution time.)
The second row corresponds to an overlapped execution of the pipeline, with one thread per stage
(4 for $B$--$E$). Compared with the runtime per stage of the serial case, 
the execution time
of all stages is increased, likely due to memory conflicts but also to the overheads of the pipeline (e.g., initial
and final latencies). 
However, the overlapped execution renders a lower total
time (617.34s for the pipelined execution vs 1,139.20s for the serial one), 
which basically agrees with that of the execution time of the slowest stage (610.39s for stage $C$), showing
a remarkable degree of overlapping.

Given the outcome of this initial experiment, 
it is natural to start by assigning a higher number of threads to the execution of stage $C$.
Consider now the third row of the table, which reports the execution time when 7 threads are employed in the computation,
4 of them for stage $C$.
This configuration reduces the execution time of stage $C$ from 610.39s to 168.32s, and the total time
to 350.04s, which was to be expected because now the bottleneck is shifted to stage $E$. As a side effect,
the time required for stage $D$ is also significantly reduced, but with no impact on the total time.

The following two configurations employ 8 threads, further reducing the total execution time, 
but with a different distribution: 1+3+1+3
and 1+4+1+2, with the first one obtaining slightly better results (312.33s vs 314.04s). 
By increasing the number of threads, while distributing them with care among the different stages, we can still
reduce the execution time up to 62 threads (remember that the remaining 2 threads are necessary for I/O). 
The best configuration in this case is 10+20+16+16 and results in a balanced execution time of 
stages $C$ and $D$, which exhibit now the longest execution times. Although, in principle, we could shift some threads from
stages $B$ and $E$ in an attempt to accelerate stage $C$, according to the experiments reported
in Table~\ref{tab:intrastage} and the top left plot of Figure~\ref{fig:intrastage}, we cannot expect much improvement
from that. Thus, any further effort to accelerate the pipeline, by shifting threads among stages, 
is useless given the bottleneck that the lack of further concurrency of stage $C$ represents at this point.
A similar analysis holds for the remaining three test cases.

\begin{table*}
  \caption{Execution time (in s.) of stages $B$--$E$ when the full pipeline is in operation for the four test cases.}
  \label{tab:full_pipeline}
  \begin{center}
    \subfloat[Test case \texttt{T1}: 100~nts, $\varepsilon=$0.1\%.]{ %
      \begin{tabular}{rrrr|c|rrrr|r}
        \toprule
        \multicolumn{4}{c|}{Threads per stage} &
        \multicolumn{1}{c|}{Total} &
        \multicolumn{4}{c|}{Time per stage} & 
        \multicolumn{1}{c}{Total} 
        \\
        \multicolumn{1}{c}{$B$}  & \multicolumn{1}{c}{$C$} & \multicolumn{1}{c}{$D$} &
        \multicolumn{1}{c|}{$E$} & 
        \multicolumn{1}{c|}{threads} &
        \multicolumn{1}{c}{$B$}  & \multicolumn{1}{c}{$C$} & \multicolumn{1}{c}{$D$} &
        \multicolumn{1}{c|}{$E$} & \multicolumn{1}{c}{time}
        \\
        \midrule
        \multicolumn{4}{c|}{Sequential version}  &  1 & 38.58 & 562.80 & 219.79 & 318.02 & 1,139.20 \\
        1  &   1  &   1  &   1  &  4 & 51.81 & 610.39 & 284.37 & 335.97 &   617.34 \\
        1  &   4  &   1  &   1  &  7 & 51.86 & 168.63 & 305.81 & 338.55 &   350.04 \\
        1  &   3  &   1  &   3  &  8 & 53.49 & 228.63 & 300.93 & 119.59 &   312.33 \\
        1  &   4  &   1  &   2  &  8 & 55.28 & 181.01 & 302.40 & 178.37 &   314.06 \\
        1  &   6  &   2  &   4  & 13 & 59.12 & 128.89 & 175.30 &  96.31 &   188.54 \\
        1  &   7  &   3  &   5  & 16 & 59.12 & 121.24 & 129.44 &  84.27 &   143.08 \\
        1  &  16  &   4  &   6  & 27 & 70.52 &  85.35 & 104.60 &  73.84 &   118.77 \\
        10  &  20  &  16  &  16  & 62 & 29.44 &  84.11 &  84.20 &  30.06 &    99.79 \\
	14  &  16  &  16  &  16  & 62 & 25.64 &  86.50 &  86.58 &  27.83 &   101.33 \\
        \bottomrule
      \end{tabular}
    }
    \\
    \subfloat[Test case \texttt{T2}: 100~nts, $\varepsilon=$1\%.]{ %
      \begin{tabular}{rrrr|c|rrrr|r}
        \toprule
        \multicolumn{4}{c|}{Threads per stage} &
        \multicolumn{1}{c|}{Total} &
        \multicolumn{4}{c|}{Time per stage} & 
        \multicolumn{1}{c}{Total} 
        \\
        \multicolumn{1}{c}{$B$}  & \multicolumn{1}{c}{$C$} & \multicolumn{1}{c}{$D$} &
        \multicolumn{1}{c|}{$E$} & 
        \multicolumn{1}{c|}{threads} &
        \multicolumn{1}{c}{$B$}  & \multicolumn{1}{c}{$C$} & \multicolumn{1}{c}{$D$} &
        \multicolumn{1}{c|}{$E$} & \multicolumn{1}{c}{time}
        \\
        \midrule
        \multicolumn{4}{c|}{Sequential version}  &  1 & 32.42 & 615.99 & 243.33 & 361.26 & 1,253.00 \\
        1  &   1  &   1  &   1  &  4 & 45.38 & 675.84 & 314.04 & 368.31 &   682.56 \\
        1  &   4  &   1  &   1  &  7 & 47.92 & 210.29 & 340.54 & 370.18 &   386.65 \\
        1  &   3  &   1  &   3  &  8 & 48.60 & 263.42 & 336.08 & 132.42 &   347.19 \\
        1  &   4  &   1  &   2  &  8 & 48.02 & 196.94 & 335.60 & 195.06 &   346.70 \\
        1  &   6  &   2  &   4  & 13 & 53.00 & 149.69 & 194.00 & 108.79 &   207.30 \\
        1  &   7  &   3  &   5  & 16 & 54.64 & 130.18 & 143.52 &  90.47 &   157.58 \\
        1  &  16  &   4  &   6  & 27 & 60.91 &  77.58 & 112.33 &  76.94 &   127.06 \\
	10  &  20  &  16  &  16  & 62 & 27.01 &  72.00 &  28.86 &  30.82 &    88.35 \\
	14  &  16  &  16  &  16  & 62 & 23.83 &  79.74 &  27.31 &  30.82 &    95.30 \\
        \bottomrule
      \end{tabular}
    }
    \\
    \subfloat[Test case \texttt{T3}: 250~nts, $\varepsilon=$0.1\%.]{ %
      \begin{tabular}{rrrr|c|rrrr|r}
        \toprule
        \multicolumn{4}{c|}{Threads per stage} &
        \multicolumn{1}{c|}{Total} &
        \multicolumn{4}{c|}{Time per stage} & 
        \multicolumn{1}{c}{Total} 
        \\
        \multicolumn{1}{c}{$B$}  & \multicolumn{1}{c}{$C$} & \multicolumn{1}{c}{$D$} &
        \multicolumn{1}{c|}{$E$} & 
        \multicolumn{1}{c|}{threads} &
        \multicolumn{1}{c}{$B$}  & \multicolumn{1}{c}{$C$} & \multicolumn{1}{c}{$D$} &
        \multicolumn{1}{c|}{$E$} & \multicolumn{1}{c}{time}
        \\
        \midrule
        \multicolumn{4}{c|}{Sequential version}  &  1 & 37.07 & 1,545.78 &   730.98 & 2,198.11 & 4,511.90 \\
        1  &   1  &   1  &   1  &  4 & 59.24 & 1,894.40 & 1,012.06 & 2,210.04 & 2,213.96 \\
        1  &   4  &   1  &   1  &  7 & 60.57 &   555.15 & 1,036.71 & 2,239.92 & 2,243.22 \\
        1  &   3  &   1  &   3  &  8 & 56.58 &   708.99 &   996.13 &   791.78 & 1,005.60 \\
        1  &   4  &   1  &   2  &  8 & 59.25 &   536.20 & 1,055.92 & 1,151.74 & 1,165.00 \\
        1  &   6  &   2  &   4  & 13 & 55.58 &   440.36 &   614.39 &   624.30 &   653.12 \\
        1  &   7  &   3  &   5  & 16 & 58.68 &   409.53 &   463.20 &   522.25 &   538.09 \\
        1  &  16  &   4  &   6  & 27 & 74.69 &   225.18 &   343.19 &   439.64 &   457.39 \\
	10  &  20  &  16  &  16  & 62 & 36.49 &   261.19 &    79.91 &   203.42 &   281.58 \\
	14  &  16  &  16  &  16  & 62 & 43.25 &   286.86 &    73.86 &   208.09 &   307.00 \\
        \bottomrule
      \end{tabular}
    }
    \\
    \subfloat[Test case \texttt{T4}: 250~nts, $\varepsilon=$1\%.]{ %
      \begin{tabular}{rrrr|c|rrrr|r}
        \toprule
        \multicolumn{4}{c|}{Threads per stage} &
        \multicolumn{1}{c|}{Total} &
        \multicolumn{4}{c|}{Time per stage} & 
        \multicolumn{1}{c}{Total} 
        \\
        \multicolumn{1}{c}{$B$}  & \multicolumn{1}{c}{$C$} & \multicolumn{1}{c}{$D$} &
        \multicolumn{1}{c|}{$E$} & 
        \multicolumn{1}{c|}{threads} &
        \multicolumn{1}{c}{$B$}  & \multicolumn{1}{c}{$C$} & \multicolumn{1}{c}{$D$} &
        \multicolumn{1}{c|}{$E$} & \multicolumn{1}{c}{time}
        \\
        \midrule
        \multicolumn{4}{c|}{Sequential version}  &  1 & 36.67 & 1,872.21 &   711.24 & 2,149.98 & 4,770.10 \\
        1  &   1  &   1  &   1  &  4 & 46.42 & 2,403.74 &   867.36 & 2,169.78 & 2,495.65 \\
        1  &   4  &   1  &   1  &  7 & 54.36 &   648.26 & 1,052.56 & 2,209.51 & 2,274.34 \\
        1  &   3  &   1  &   3  &  8 & 51.54 &   844.20 &   988.80 &   809.24 &   999.75 \\
        1  &   4  &   1  &   2  &  8 & 51.76 &   633.85 & 1,024.40 & 1,125.76 & 1,136.65 \\
        1  &   6  &   2  &   4  & 13 & 51.75 &   471.66 &   595.90 &   653.56 &   671.46 \\
        1  &   7  &   3  &   5  & 16 & 54.85 &   416.75 &   456.66 &   508.98 &   526.43 \\
        1  &  16  &   4  &   6  & 27 & 65.80 &   256.04 &   341.97 &   428.24 &   446.64 \\
	10  &  20  &  16  &  16  & 62 & 37.62 &   285.48 &    82.72 &   211.68 &   306.29 \\
	14  &  16  &  16  &  16  & 62 & 40.06 &   313.45 &    73.51 &   216.00 &   333.45 \\
        \bottomrule
      \end{tabular}
    }
  \end{center}
\end{table*}

Figure~\ref{fig:full_pipeline} shows the speed-up of the full pipeline
(serial time divided by parallel time) for up to 62+2 threads.
These results show performance improvements in the range of 7.9--9.0 for 16+2 threads,
9.6--10.6 for 27+2 threads, 
and 11.4--16.0 using 62+2 threads.
For benchmarks {\tt T1}--{\tt T2}, for example, this represents a reduction of
the mapping process from around 19--20 minutes
to about 1.5 minutes.  
For benchmarks {\tt T3}--{\tt T4}, on the other hand, the reduction is from approximately 1 hour and 15--20 minutes to
only about 5 minutes.

\subsection{Comparison with TopHat~2}

We have compared \rnaseq{} to the most extensively used RNA-seq
mapper, TopHat~2, combined with both Bowtie and Bowtie~2 using the
four test cases, on a platform with 2 Intel Xeon E5645 processors at
2.40 GHz (six cores per processor), and 48 Gbytes of RAM (ccNUMA).  To
obtain the following results, all the RNA-seq software, including our
\rnaseq{}, was executed with the default input parameters, using all
cores of the platform.  For that purpose, we set the flag {\tt p}=12
for TopHat~2 while we employed 1+6+2+3 threads for the pipelined
aligner.  The results in Table~\ref{tab:comparison} show that
\rnaseq{} is consistently over six times faster than TopHat~2 combined
with Bowtie~2. In addition, the sensitivity of the proposed aligner,
measured as the percentage of reads that were correctly mapped,
clearly outperforms that of TopHat~2 with both Bowtie and
Bowtie~2. When the read length grows, this difference in sensitivity
is even more acute, as \rnaseq{} increases its sensitivity while that
of TopHat~2 drops.  Similar results were obtained when the aligners
were applied to real datasets.

\begin{table*}
  \caption{Comparison of sensitivity (in \%) and runtime (in minutes)
    of three RNA-seq aligners. In those cases labelled as ``NA'', the
    program was stopped after three days in execution, with no results
    produced at that point.}
  \label{tab:comparison}
  \centering
  \begin{tabular}{r|rr|rr|rr}
    \toprule
    &
    \multicolumn{2}{c|}{\rnaseq{}} &
    \multicolumn{2}{c|}{TopHat~2 + Bowtie} &
    \multicolumn{2}{c}{TopHat~2 + Bowtie~2}
    \\ 
    & Sensitivity & Time  & Sensitivity &  Time & Sensitivity & Time  \\ \midrule
    {\tt T1} &       97.15 &  2.41 &       62.47 & 23.71 &     62.09 & 20.13 \\
    {\tt T2} &       95.88 &  2.66 &       46.78 & 21.65 &     47.64 & 25.05 \\
    {\tt T3} &       97.32 & 14.76 &          NA & NA    &        NA &    NA \\
    {\tt T4} &       97.15 & 14.52 &          NA & NA    &        NA &    NA \\
    \bottomrule
  \end{tabular}
\end{table*}


\section{Conclusions}\label{sec:conclusions}

We have introduced a pipeline for RNA sequencing, the \rnaseq{}, that efficiently accommodates 
variable levels of hardware concurrency in current multicore technology, 
enabling fast mapping onto a reference genome, with a cost that linearly depends 
on the number and length of the RNA fragments.
Our solution leverages a well-known principle, that of ``{\em making the common case fast}'', 
to apply a variant of BWT in order to map those reads with at most 1 EID in a short period of time. 
(Note that the reliability of current NGS technology ensures that this part constitutes a
large fraction of the total.)
After this initial stage, mapping failures are expected to be mostly due to reads with more than 1 EID or,
alternatively, reads that span over two or more exons. To tackle both cases, we proceed  
by dividing the reads into a collection of short seeds, 
which are next mapped using the fast BWT (again), yielding a collection of 
CALs in the reference genome. This information is finally passed to the accurate SWA that, 
under these conditions, turns most of the previous failures into successful mappings at a low cost.

Experiments on a server with a high number of cores reveal the parallel efficiency of the \rnaseq{} pipeline,
which is ultimately constrained by the serial performance of the expensive Region seeker.
Future work will therefore aim at ameliorating part of the bottleneck imposed by this stage, 
possibly by further subdividing it
into new (sub-)stages or shifting part of its cost towards neighboring stages.



\section*{Acknowledgments}

The researchers from the Universidad Jaume~I (UJI) were supported by
project TIN2011-23283 and FEDER.

\bibliographystyle{IEEEtran}

\bibliography{enrique,hector}

\end{document}